# Second-order phase transition at high-pressure in $A^4B^6$ crystals

**F.M. Hashimzade, D.A. Huseinova, Z.A. Jahangirli, B.H. Mehdiyev**

Institute of Physics, National Academy of Sciences of Azerbaijan, AZ 1143, Baku, Azerbaijan

**Abstract.** The existence of second-order structural phase transition in the SnS at a pressure of 16 GPa has been proved theoretically. The calculation is performed using the plane-wave pseudopotential approach to density-functional theory within the local-density approximation (LDA) with the help of the ABINIT software package. The abrupt change in volume compression with unit cell volume continuous change of the crystal is the clear evidence of the second-order phase transition. It is shown that the phase transition is caused by the softening of the low-frequency fully symmetric interlayer mode with increasing pressure. As a result, displacement type phase transition (PT) take place with the change of translational symmetry of the crystal from the simple orthorhombic to the base-centered orthorhombic ($P_{cmn}$ ($D_{2h}^{16}$) $\rightarrow$ $C_{mcm}$ ($D_{2h}^{16}$)).

## 1. Introduction

In recent years, great efforts are aimed at creating photovoltaic devices from non-toxic materials with a simple, low cost production technology. In this regard, very promising were semiconductor compounds of group $A^4 B^6$. Preliminary solar cell devices incorporating SnSe nanocrystals into a specific matrix demonstrate a significant enhancement in quantum efficiency and short-circuit current density, suggesting that this earth-abundant material could be a valuable component in future photovoltaic devices[ 1]. For example, SnS used as absorbing layer in thin-film solar energy converters [2-5], as well as the photoconductors [6], semiconductor sensors [7] and mikro- batteries[8].

Much of modern microelectronics is based on the use of thin films grown on various substrates. A mismatch between the lattice parameters of the film and those of the substrate results in a compression or a tension in the film. Furthermore, the difference between the thermal expansion coefficient of the film and that of the substrate leads to biaxial deformations. The structural parameters of the lattice and the electronic properties of the crystals change substantially under the resulting pressure, and this should be taken into account when constructing various devices. It is, therefore, important to study the effect of pressure on the structural, elastic and electronic parameters of the compounds used in such devices.

A number of theoretical studies of the effect of hydrostatic pressure on the structural parameters and the phonon spectra of the A4B6 compounds with layered structure have been published in recent years.

Hsueh et al. [9,10,11] calculated structural properties and phonon spectra of GeS under pressure from the first principles, in the plane-wave basis, using the pseudopotential approach to density functional theory in the local density approximation (LDA). The authors concluded that the density functional theory adequately describes not only the equilibrium structure and the vibrational properties of the strongly anisotropic GeS compound, but also effectively predicts the details of pressure-induced changes in the lattice parameters and atomic positions in the unit cell. The authors demonstrated theoretically and observed in the experiments that there is no phase transition (PT) in this compound at the pressures of up to 9.4 GPa. In the earlier experimental studies of the A4B6 compounds with orthorhombic lattice no PT have been detected at the pressures. [12-14].

However authors [15] are found the PT in SnS at the pressure 18GPa . In the theoretical work [16] authors support this result.

## 2. Crystal Structure

It is known that the following four compounds from $A^4B^6$ group (GeS, GeSe, SnS, SnSe) have an orthorombic lattice structure. The unit cell of the crystal contains eight atoms arranged in two layers, each consisting of four atomic planes in sequence:the cation-anion-anion-cation. The intralayer bonds are predominantly of a covalent character, whereas the bond between the layers is weak and, supposedly, of the van der Waals type. The positions of the atoms, in fractional coordinates, in the unit cell are follows: both types of atoms occupy positions (4c) with the coordinates ±(x,1/4,z), ±(1/2-x, 3/4,1/2+z). The crystal structure of above mentioned semiconducting compounds is shown in Fig.1. Space group symmetry is $P_{cmn}$ ($D_{2h}^{16}$) [15]. The structure parameters are shown in Table1

Table 1. Structure parameters GeS, GeSe, SnS, SnSe[15]

| Compound / Lattice parameters | GeS | GeSe | SnS | SnSe |
|---|---|---|---|---|
| a(Å) | 4.299(2) | 4.388(4) | 4.334(1) | 4.445(1) |
| c(Å) | 10.481(4) | 10.825(9) | 11.200(2) | 11.501(2) |
| b(Å) | 3.646(2) | 3.833(4) | 3.987((1) | 4.153(1) |

| Atom | Ge | S | Ge | Se | Sn | S | Sn | Se |
|---|---|---|---|---|---|---|---|---|
| x | 0.1277(1) | 0.5023(4) | 0.1115(1) | 0.5020(2) | 0.1198(2) | 0.4793(8) | 0.1035(3) | 0.4819(4) |
| z | 0.1221(1) | 0.8495(1) | 0.1211(1) | 0.8534(1) | 0.1194(1) | 0.8508(3) | 0.1185(1) | 0.8548 |

## 2. Calculation method

We have employed the method of density functional in the local density approximation, using a pseudopotential and the plane-wave expansion of the wave function, in the ABINIT software package [16]. The exchange-correlation interaction was described in the local density approximation, as in [17]. For the pseudopotentials we used the norm-conserving pseudopotentials of Hartwigsen-Goedecker-Hutter (HGH) [18]. In the expansion of the wave function we used plain waves with the maximal kinetic energy of up to 0.2 fJ, which ensures a good convergence of the total energy. An integration over the Brillouin zone (BZ) has been carried out using the ($4 \times 4 \times 4$) Monkhorst-Pack grid [19], with the shift by (0.5, 0.5, 0.5) from the origin. The lattice parameters and the equilibrium locations of atoms in a unit cell were determined by the minimization of Hellmann-Feynman forces. The minimization procedure was iterated until the modules of the forces fall below 8 fN for each given level of pressure.

The lattice dynamics was calculated in the density functional perturbation theory (DFPT) framework [16]. In the DFPT the static linear response to phonon distortions is determined by the electronic properties of the ground state. Using DFPT, the dynamical matrices can be obtained for any point of the BZ.

We computed the dynamic matrices and the phonon frequencies for certain point of the BZ using the elements of the space group symmetry. After that, we calculated the interatomic force constants using Fourier transform[1]. Finally, an inverse Fourier transform allows to compute the dynamic matrix for any wave vector in the BZ, and, therefore, to compute all phonon frequencies. For the phonon density of states we used the 40x40x40 –partition of the BZ [20-22].

## A. GeS

The theoretical calculation of the dependence of the relative change in the crystal volume with pressure is shown in Figure 2, along with the experimental data from [9] and [12]. Figure 3 shows the pressure dependence of the bulk modulus of the GeS crystal obtained from the computed values of the components of the tensor of compliance coefficients at different pressures. Figure 4 shows the change in the elastic moduli of GeS with pressure. The calculated pressure dependence of the lattice constants together with the data from [11] are shown in Figure 5, and Figures 6(a,b) illustrate the calculations of the pressure dependence for the internal structural parameters. The experimental baric dependence of the low-frequency fully symmetric Ag phonon mode frequency [9] and our calculation results are shown in Fig.7.

## B. GeSe

The theoretical calculation of the dependence of the crystal volume GeSe with pressure is shown in Fig. 8, together with the experimental data from [14]. Fig. 9 shows the pressure dependence of the bulk modulus of the GeSe crystal obtained from the computed values of the components of the tensor of compliance coefficients at different pressures. Figure 10 shows the change in the elastic moduli of GeSe with pressure.

The calculated pressure dependence of the lattice constants GeSe together with the data from [10] are shown in Figure 11. Figures 12(a,b) illustrate the pressure dependence of the internal structural parameters of GeSe.

Fig. 13 show the calculate baric dependence of the Ag phonon mode frequency of GeSe.

## C. SnS

Theoretically calculated dependence of the lattice constants and crystal volume with pressure is shown in Figure 14. Fig. 15 shows the pressure dependence of the linear and volume compressibility of the crystal SnS obtained from the calculated values components of the compliance tensor at different pressures. Fig.16 illustrate the pressure dependences of the elastic constants of SnS. Figure 17 illustrate the pressure dependence of the internal structural parameters of SnS.

---

[1] We used the ANADDB routine in the ABINIT software package for this calculation.

Fig.18 show the pressure dependence of the low-frequency fully symmetric mode $A_g$ of SnS.

## D. SnSe

The calculated pressure dependence of the dimensionless lattice constants are shown in Fig 19. Theoretically calculated dependence of the crystal volume with pressure is shown in Fig. 20. Fig.21shows the calculated pressure dependence of the internal structural parameters. Fig.22 and 23 shows the pressure dependence of the linear compressibility and bulk modulus of the crystal SnSe obtained from the calculated values components of the compliance tensor at different pressures..

Fig.24 shows the calculated pressure dependence of the low-frequency fully symmetric mode $A_g$ .

## 4. Discussion of results

As we see from Fig. 2 and Fig.4 for GeS, Fig.8 and Fig.11 for GeSe, Fig.14 for SnS , Fig. 19 and Fig.20 for SnSe, both the experimental data and the calculations show that the lattice parameters and volume of the crystals GeS, GeSe, SnS and SnSe changes continuously as pressure increases. On the other hand, the linear compressibility and bulk modulus for the all mentioned crystals have discontinuity in the change of the bulk modulus and the linear compressibility at the pressure of 35.4, 28.5, 16, 10 GPa for GeS,GeSe,SnS and SnSe , correspondingly. The continuous change of the lattice parameters and unit cell volume of the crystal with pressure and a discontinuity in the change of the bulk modulus and the linear compressibility with the pressure suggests a second-order structural phase transition (PT) in these crystals according to Landau theory of the second- order PT [25].

The authors of [12] failed to detect the second-order PT, because they have not examined the relevant physical parameters corresponding to the second derivatives of the thermodynamic potential, such as compressibility, which necessarily needed to detect second-order phase transition.

The convergence of the structural parameter $x_{Ge}$ ($x_{Sn}$) to zero and the convergence of the structural parameter $x_S$ ($x_{Se}$) to 1/2 (see Fig.6, Fig.12, Fig.17, Fig.21) indicate the change in the symmetry of the crystals from the simple orthorhombic to base-centered orthorombic structure ($P_{nma}$ ( $D_{2h}^{16}$ ) $\rightarrow$ $C_{mcm}$ ( $D_{2h}^{17}$ )).

Figures 7,13, 18, 24 show the softening of the low-frequency fully symmetric mode $A_g$ as the pressure approaches to the pressure of 35.4, 28.5 , 16, 10 GPa for GeS, GeSe, SnS and SnSe , correspondingly. Therefore we can conclude that the change in the symmetry is the consequence of a PT of the displacement type induced by softening of the fully-symmetrical low-frequency interlayer mode.

The phase transition induced by pressure at 18.1 GPa, discovered by the authors of [13] for SnS, as it appears, related to the structural phase transition described here. Indeed, if the translation vectors base-centred orthorhombic lattice are transformed into a monoclinic one, we obtain the following parameters: $a_m$ = 10.945Å , $b_m$ = 3.835Å , $c_m$ = 7.511Å , $\beta$ = 110.73. These are close to the parameters given in [14].

**5. Conclusion**

In this paper have been investigated the vibrational spectrum of A4B6 crystal in the framework of the density functional theory within the local-density approximation. We determined the relative change in volume with pressure, the pressure dependence of the bulk modulus of compressibility, the pressure dependences of the elastic moduli, and the pressure dependence of the internal parameters of A4B6 crystal. We established the existence of the second-order PT near the pressure of 35, 28,16,10 GPa, respectively. This second-order structure phase transition is displacement type and is induced by the softening of the low-frequency fully symmetric interlayer shear vibration mode $A_g$ ($P_{cmn}$ ($D_{2h}^{16}$) $\rightarrow$ $C_{mcm}$ ($D_{2h}^{16}$)).

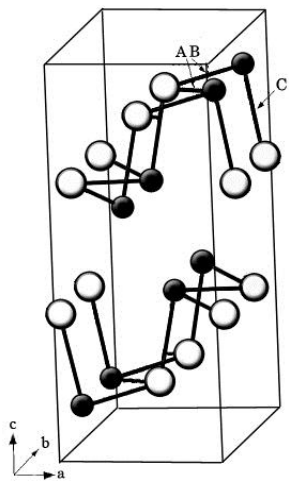

Fig.1. GeS structure. Ge and S are represented as small and big spheres, respectively. The unit cell contains two double layers and each double layer is formed by covalently bonded Ge-S pairs in which atoms are threefold coordinated inside the double layer. For α-GeS, there are two distinct bond lengths.the bond nearly parallel to the crystallographic *c* axis as the *C* bond and the *AB* bond in the *a*-b plane of the layers. (From Ref.[9]).

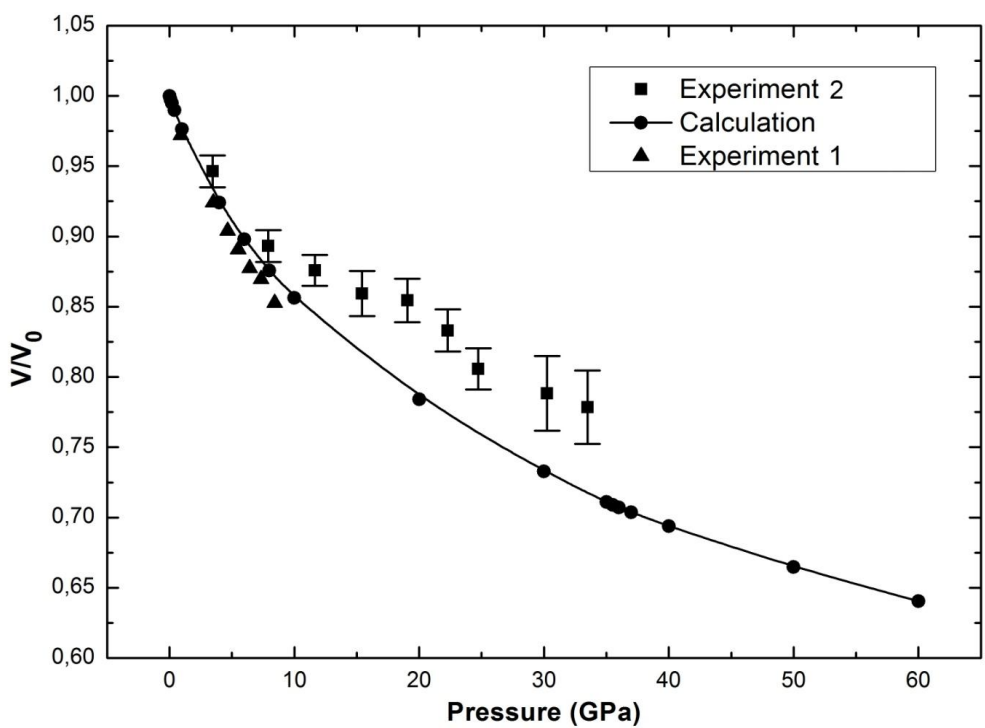

Fig.2. The normalised unit cell volume $V/V_0$ ($V_0$ is the equilibrium unit cell volume at ambient pressure) as a function of hydrostatic pressure. The triangles and squares correspond to the experimental data extracted from Ref. [1] and Ref. [3], respectively. The solid line corresponds to the best fit to the calculated values.

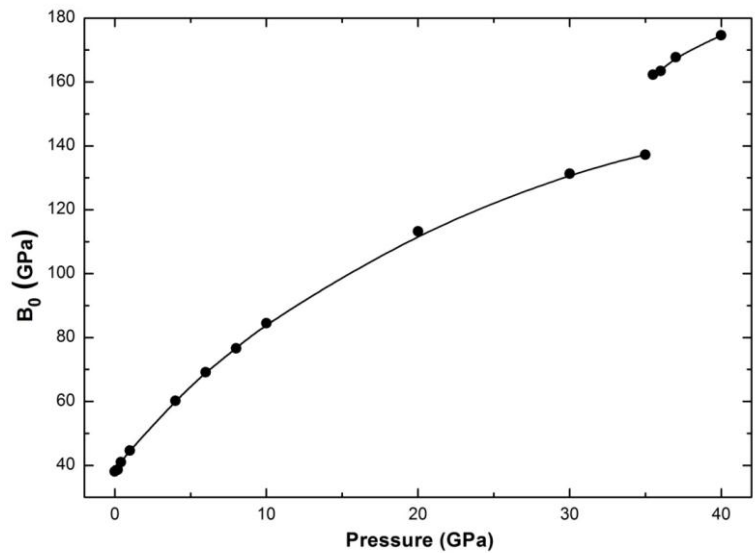

Fig.3. The theoretical pressure dependence of the bulk modulus of the GeS crystal. The solid line is the best fit of the calculated points.

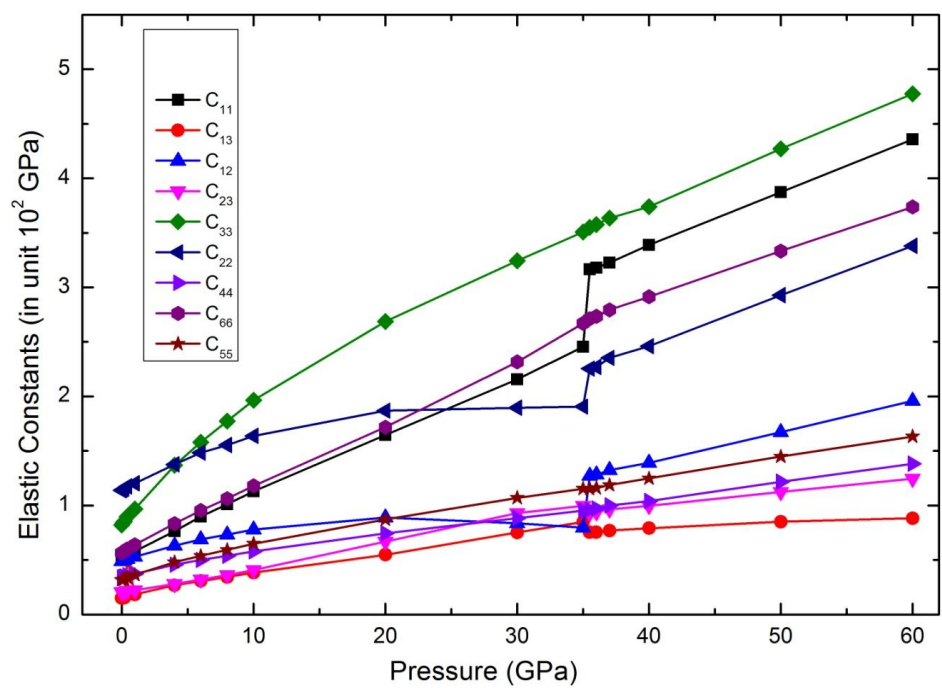

Fig.4. The theoretical pressure dependence of the elastic moduli of GeS.

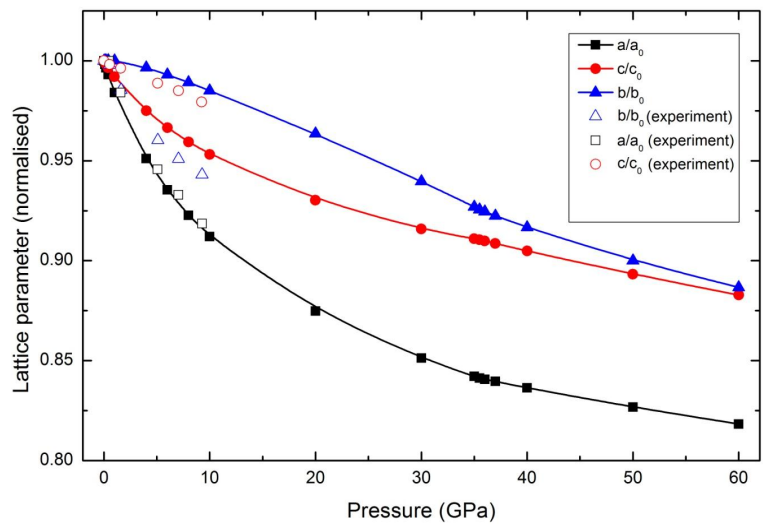

Fig.5. The normalized lattice parameters of GeS as a function of hydrostatic pressure. Solid symbols correspond to the calculated values and open symbols correspond to experimental data extracted from Ref.[2].

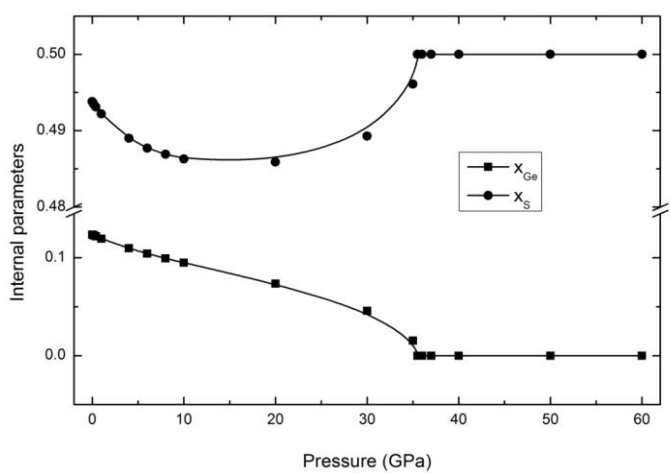

Fig. 6(a)

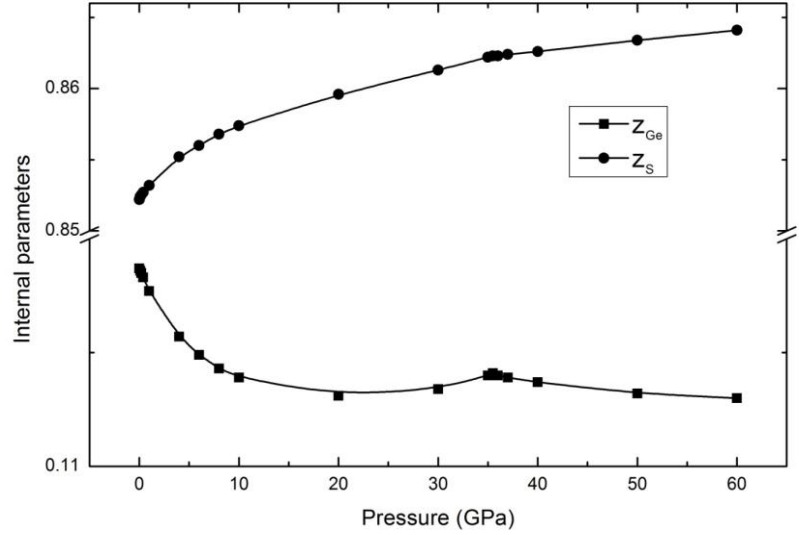

Fig. 6(b)

Fig. 6. The calculated pressure dependence of the internal structural parameters of GeS.

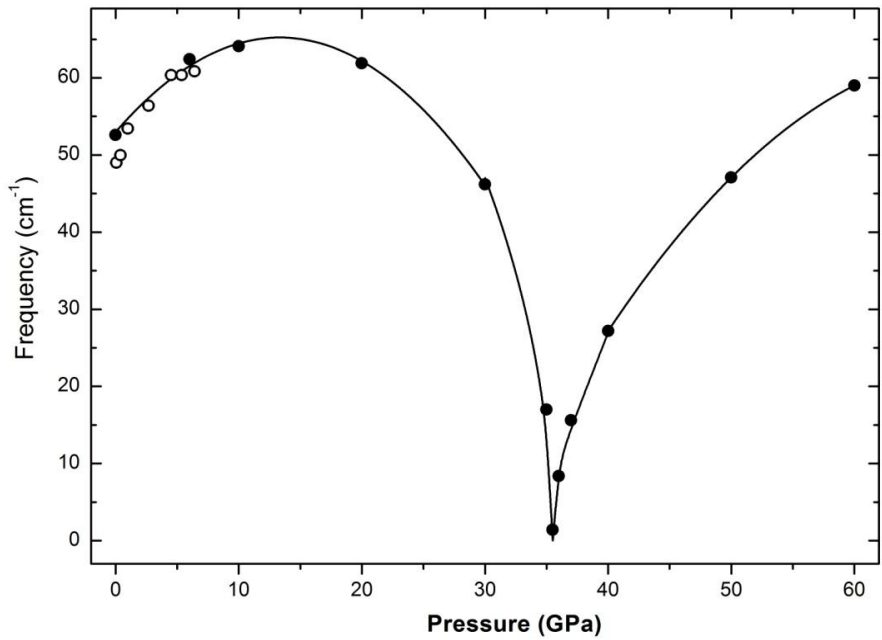

Fig. 7. The hydrostatic pressure dependence of the $A_g$ shear phonon mode. Solid symbols correspond to the calculated values and open symbols correspond to experimental data extracted from Ref.[1].

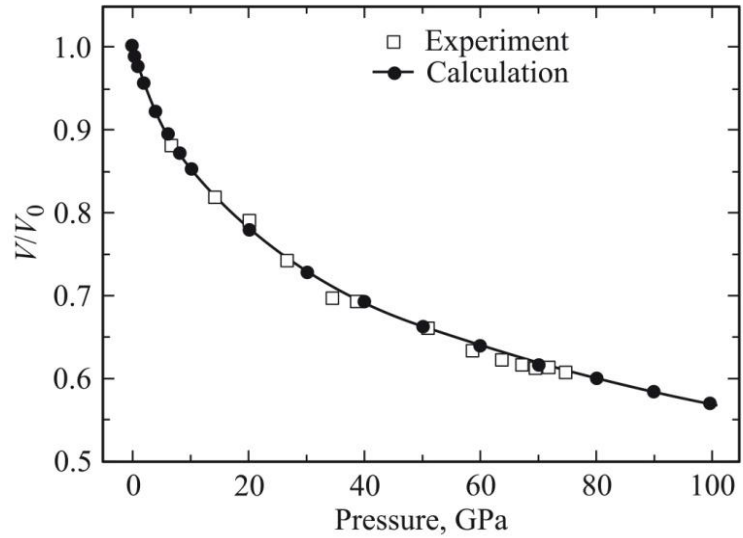

Fig.8. The normalized crystal volume of GeSe as a function of hydrostatic pressure.

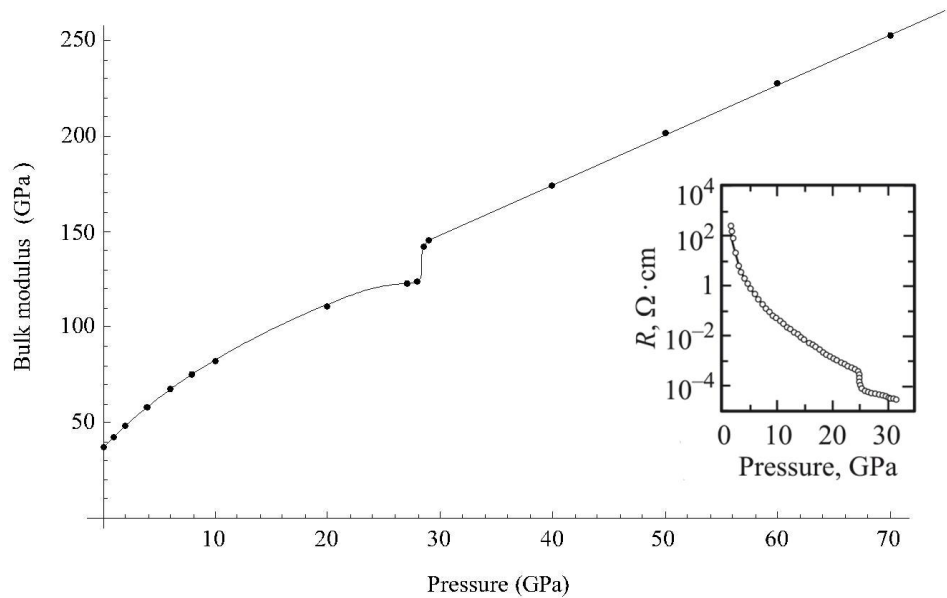

Fig.9 Bulk modulus GeSe . The inset of Fig. shows the plot of electrical resistivity of GeSe as a function of pressure extracted from Ref.[14]. The sharp change in the bulk modulus near 29 GPa correlates with a sharp drop in the resistance at 25 GPa.

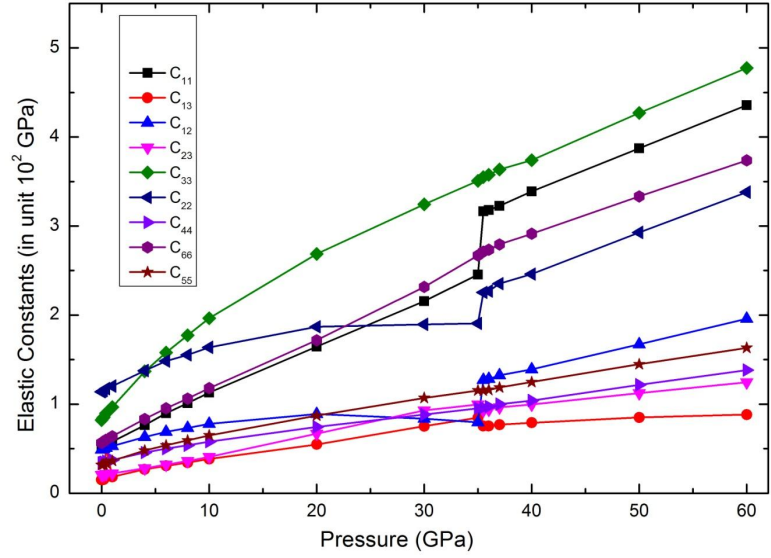

Fig.10. The pressure dependence of the elastic constants of GeSe.

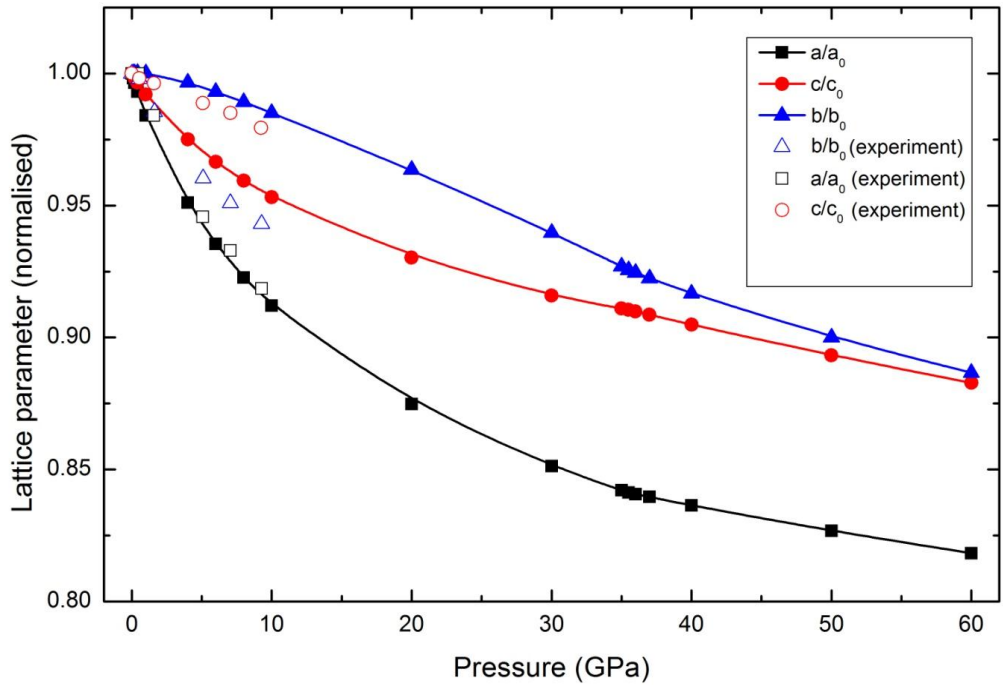

Fig.11. The normalized lattice parameters of GeSe as a function of hydrostatic pressure.

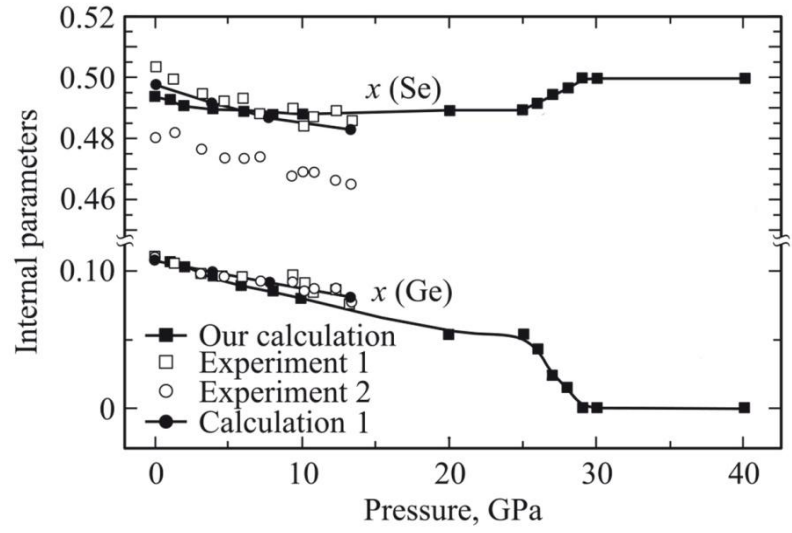

Fig. 12a. Pressure dependence of internal parameters GeSe.

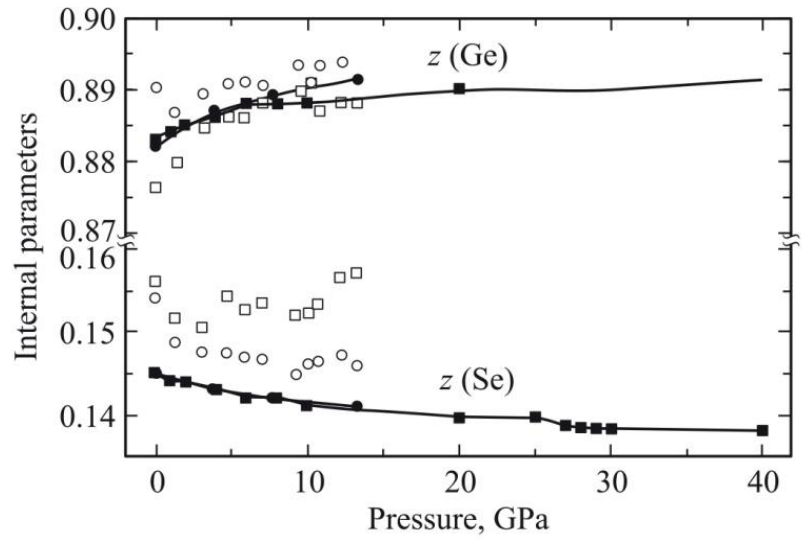

Fig.12b. Internal parameters GeSe

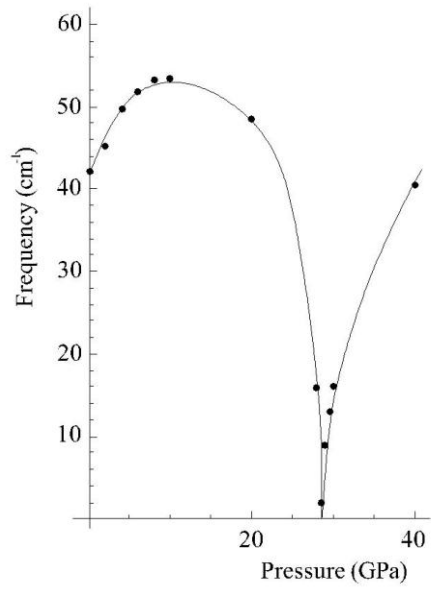

Fig.13.The hydrostatic pressure dependence of the Ag shear phonon mode of GeSe.

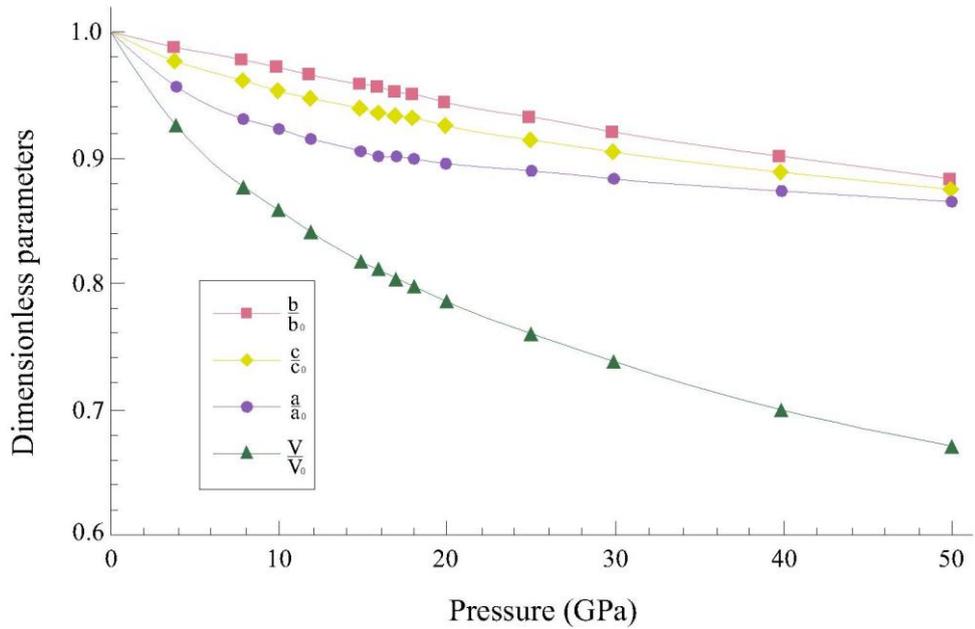

Fig.14. The lattice parameters and volume of SnS.

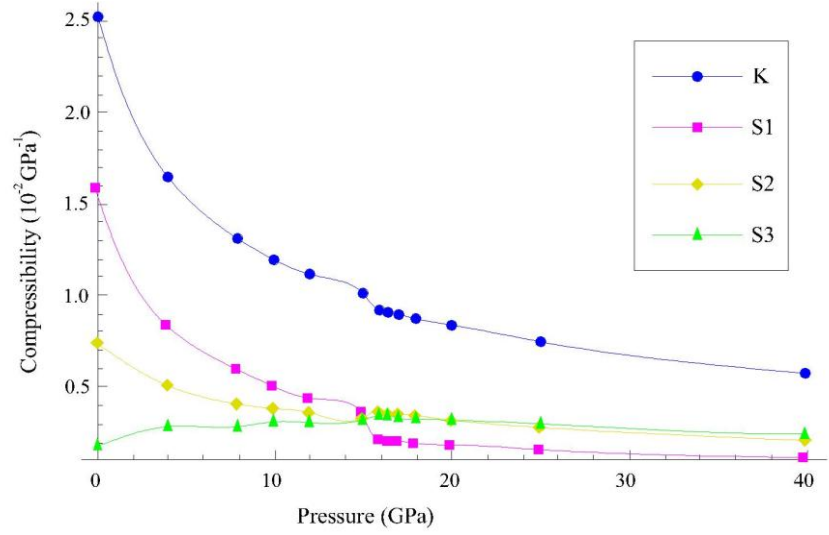

Fig.15.Linear compressibility and bulk modulus of SnS.

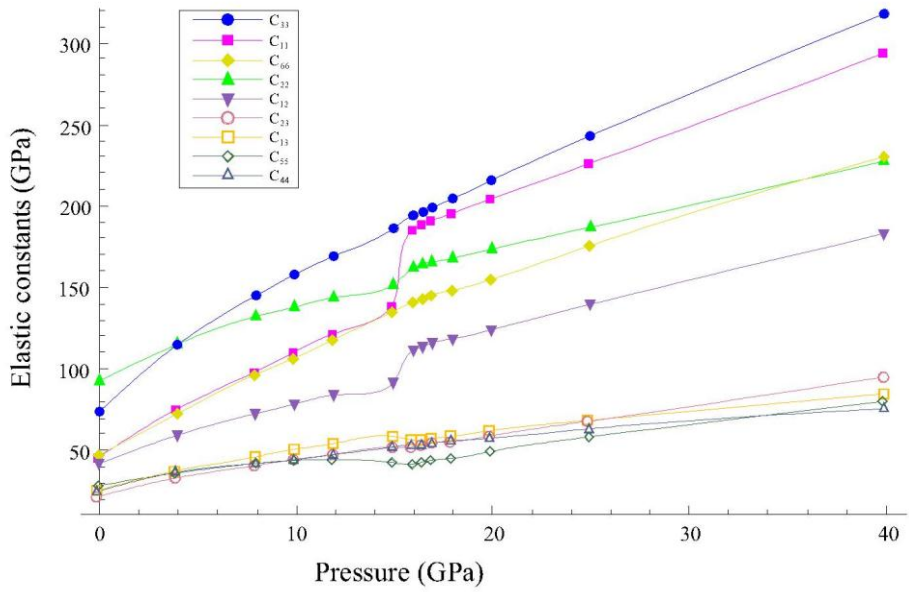

Fig.16. Elastic constants of SnS.

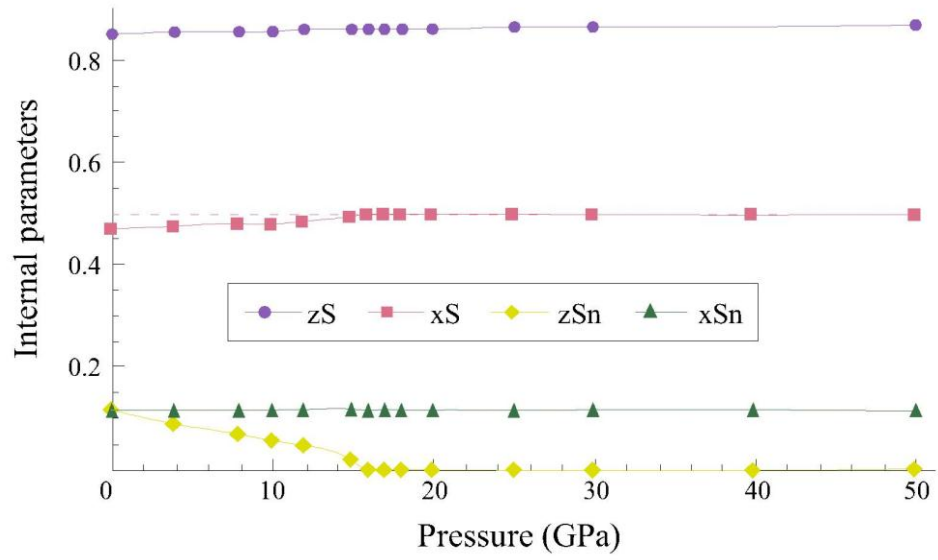

Fig.17. Internal parameters of SnS.

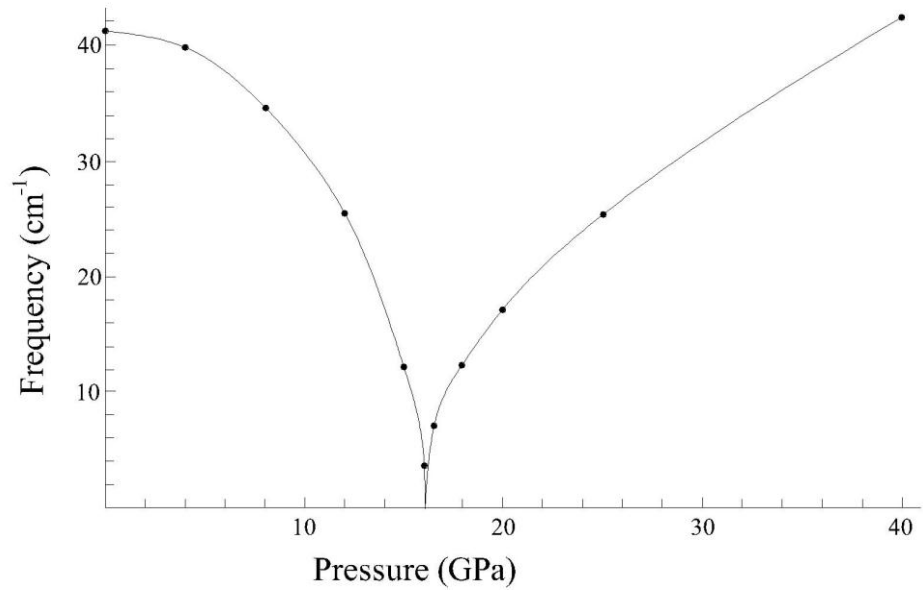

Fig.18. The hydrostatic pressure dependence of the Ag shear phonon mode of SnS.

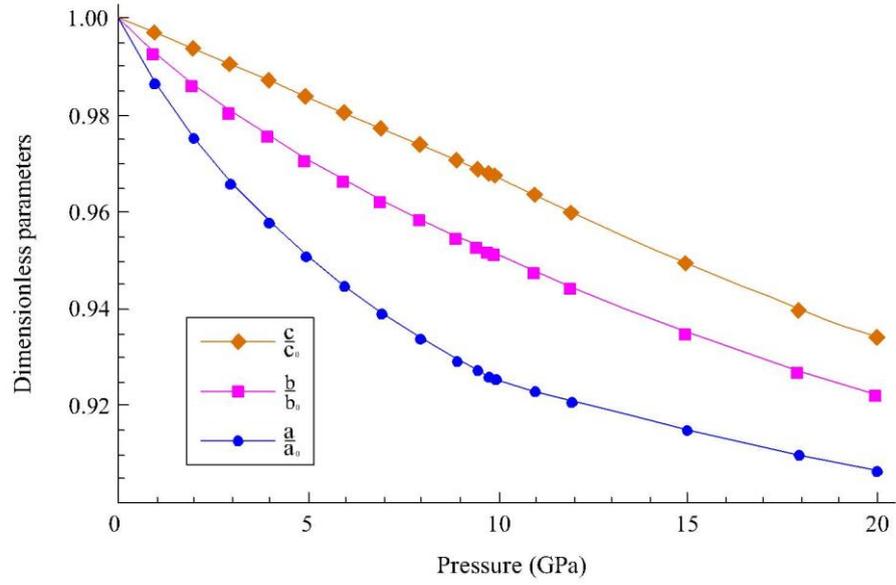

Fig.19. The normalized lattice parameters SnSe as a function of hydrostatic pressure.

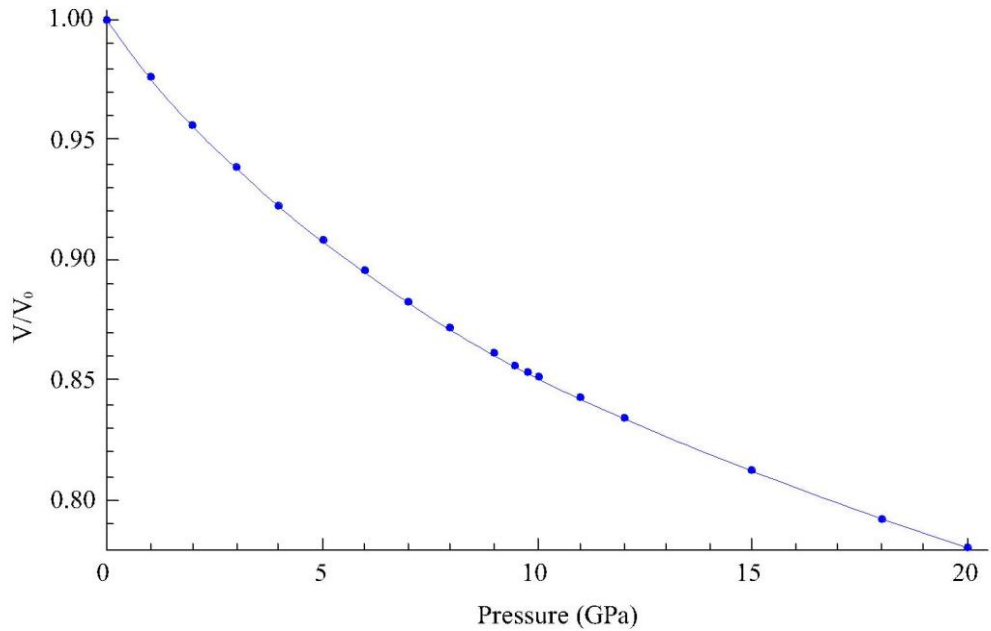

Fig.20. The normalized crystal volume of SnSe as a function of hydrostatic pressure.

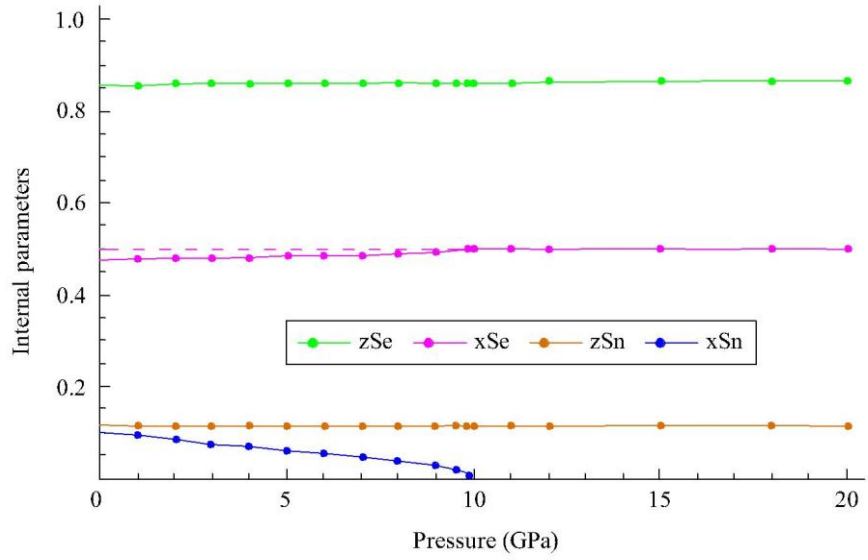

Fig.21. The calculated pressure dependence of the internal structural parameters of SnSe.

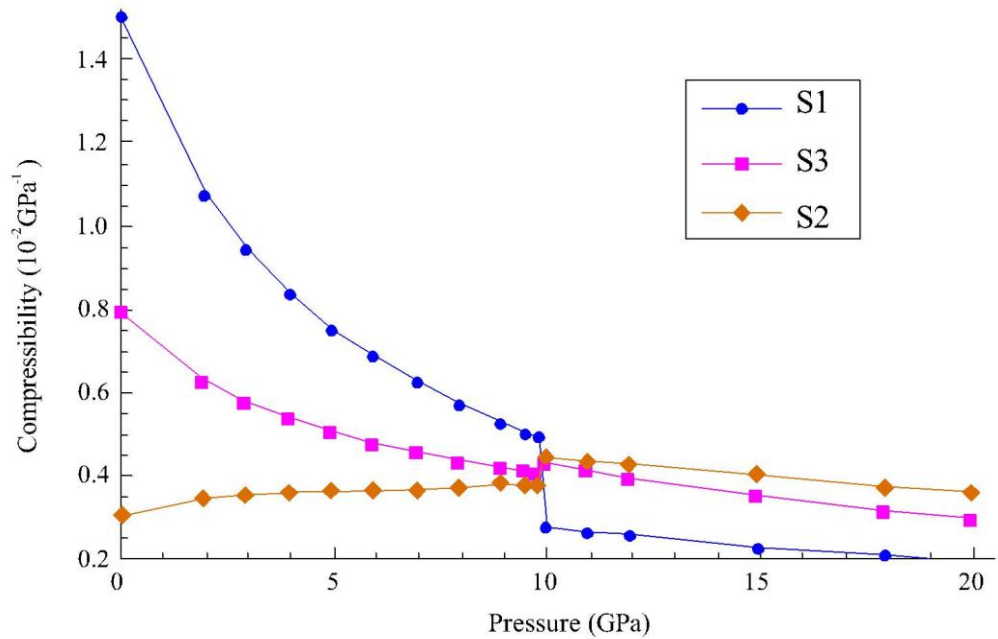

Fig.22. The pressure dependence of the linear compressibility of the crystal SnSe.

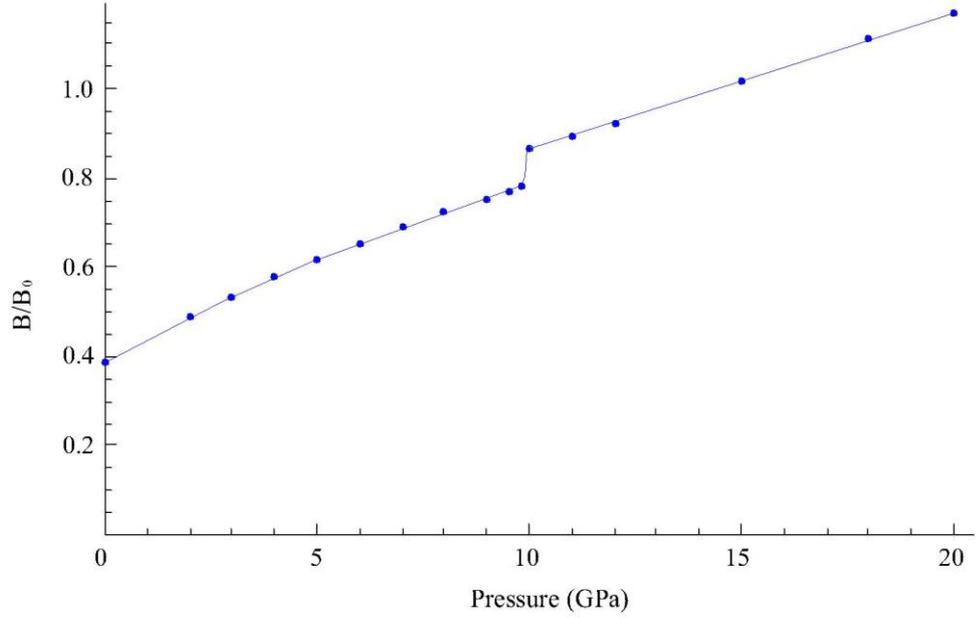

Fig.23. The theoretical pressure dependence of the bulk modulus of the SnSe crystal.

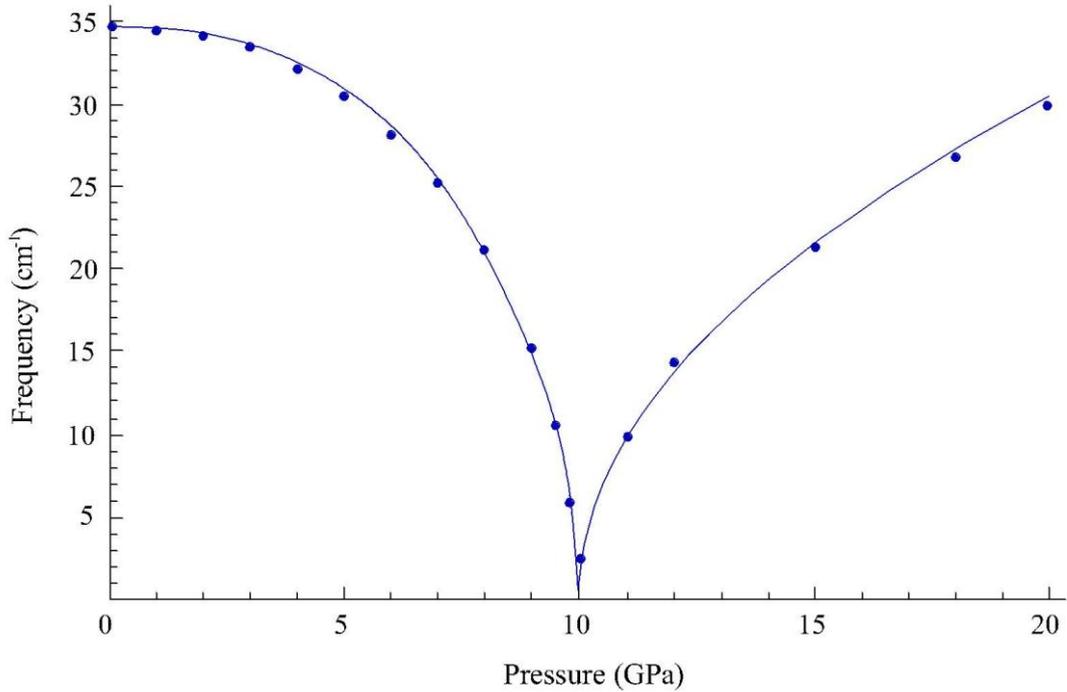

Fig.24. The hydrostatic pressure dependence of the Ag shear phonon mode of SnSe.